\documentclass[structabstract]{aa}  
\usepackage{graphicx}
\usepackage{natbib}
\bibpunct{(}{)}{;}{a}{}{,}
\usepackage{txfonts}
\begin{document}
   \title{A new perspective on GCRT J1745-3009}

   \author{H. Spreeuw
          \inst{1}, B. Scheers \inst{1}, R. Braun\inst{2,6}, R.A.M.J. Wijers\inst{1}, J.C.A. Miller-Jones\inst{3},
          B.W. Stappers\inst{4,6} \and R.P. Fender\inst{5}}

   \institute{\inst{1} Astronomical Institute "Anton Pannekoek", University of Amsterdam,
              Kruislaan 403, 1098 SJ  Amsterdam, The Netherlands\\
              \email{[j.n.spreeuw,l.h.a.scheers,r.a.m.j.wijers]@uva.nl}\\
             \inst{2} Australia Telescope National Facility, CSIRO, PO Box 76, Epping NSW 1710, Australia\\
              \email{robert.braun@csiro.au}\\
             \inst{3} Jansky Fellow, National Radio Astronomy Observatory, 520 Edgemont Road, Charlottesville, VA 22903, USA\\
              \email{jmiller@nrao.edu}\\
             \inst{4} University of Manchester, Jodrell Bank Observatory, Macclesfield, Cheshire SK11 9DL, UK\\
              \email{ben.stappers@manchester.ac.uk}\\
             \inst{5} School of Physics and Astronomy, University of Southampton, Highfield, Southampton, SO17 1BJ, UK\\
              \email{rpf@phys.soton.ac.uk}\\
             \inst{6} ASTRON, PO Box 2, 7990 AA Dwingeloo, The Netherlands.}

   \date{Received September 15, 1996; accepted March 16, 1997}

 
  \abstract
   {Reports on a transient source about 1.25$\degr$
  south of the Galactic Centre motivated these follow-up observations
   with the WSRT and the reinvestigation of archival VLA data. The source GCRT J1745-3009 was detected during a 2002
  Galactic Centre monitoring programme  
   with the VLA at 92 cm by five powerful 10-min bursts with a 77-min recurrence while
   apparently lacking any interburst emission.}
   {The WSRT observations were performed and archival VLA data were reduced to re-detect GCRT J1745-3009 at
  different epochs and frequencies, to constrain its distance and
    determine its nature. We attempted to extract a  more accurate lightcurve from the discovery dataset of GCRT J1745-3009 in order to rule out some of the models that have been suggested. We also investigated transient behaviour of a nearby source.}
  {The WSRT data were taken in the 
  "maxi-short" configuration, using 10 s integrations, on 2005 March 24
  at 92 cm and on 2005 May 14/15 at 21 cm. Five of the six VLA observations we reduced are the oldest of this field in this band.} 
  {GCRT J1745-3009 was not redetected. 
   With the WSRT we reached an rms sensitivity of 0.21 mJy
  $\mathrm {beam^{-1}}$ at 21 cm and 3.7 mJy $\mathrm {beam^{-1}}$ at 92 cm.  Reanalysis of the discovery observation data resulted in a more accurate and more complete lightcurve. The five bursts appear to have the same shape: a steep rise, a more gradual brightening and a steep decay. We found variations in burst duration of order $\simeq3\%$. We improved the accuracy of the recurrence period of the bursts by an order of magnitude: $77.012\pm0.021$ min. We found no evidence of aperiodicity. We derived a very steep spectral index: $\alpha=-6.5\pm3.4$. We improved the $5\sigma$ upper limits for interburst emission and fractional circular polarization to 31 mJy $\mathrm{beam^{-1}}$ and $8\%$, respectively. Transient behaviour of a nearby source could not be established.}
  {Models that predict symmetric bursts can be ruled out, but rotating systems are favoured, because their periodicity is precise. Scattering constraints imply that GCRT J1745-3009 cannot be located far beyond the GC. If this source is an incoherent emitter and not moving at a relativistic velocity, it must be closer than 14 pc.}

   \keywords{GCRT J1745-3009 -- low frequency
                radio transients
               }

   \titlerunning{Observations of GCRT J1745-3009}
   \authorrunning{Spreeuw et al.}

   \maketitle
%

\section{Introduction}

  Reports on a peculiar radio transient, GCRT J1745-3009, about
  1.25$\degr$ south of the Galactic Centre \citep{Hyman1,Hyman2,Hyman3} and the suggestion
  that this may be the prototype of a new class of particularly bright,
  coherently emitting radio transients have led to speculation about its nature. In particular, the 77 minute recurrence of the Jy level bursts was attributed to a period of rotation \citep{ZhangGil}, revolution \citep{Turolla} and precession \citep{ZhuXu}.
A nulling pulsar and an 'X-ray quiet, radio-loud' X-ray binary have also been suggested \citep{Kulk2005} as well as an exoplanet and a flaring brown dwarf \citep{Hyman1}. The discovery has led
  to follow-up observations and re-examination of archival data at both 92 cm
  and other bands. Those did not reveal a source \citep{ZhuXu,Hyman1,Hyman2}, with two exceptions \citep{Hyman2,Hyman3}. Both of the redetections were single bursts, possibly due to the sparse sampling of these observations. The first redetection was possibly the decaying part of a bright (0.5 Jy level) burst that was detected at the first two minutes of a ten minute scan. The second redetection was a faint short ($\simeq 2$ minute) burst that was completely covered by the observation. The average flux density during the burst was only  $57.9\pm6.6$ mJy/beam. This redetection also showed evidence for a very steep spectral index ($\alpha=-13.5\pm3.0$). \\
In summary, the source has
  only been detected at three epochs, separated by less than 18 months, all at 92 cm,
  while the source was not detected in this band at 33 epochs over a period of
  more than 16 years \citep[see][ table~1]{Hyman2} nor in any other band, ever. 
  We observed the field containing GCRT J1745-3009 using eight 10-MHz IFs in the 92 cm band because its possible association with the supernova remnant G359.1-0.5 would mean that this source is about as far as the Galactic Center. That, in turn, implies
  a substantial dispersion measure (DM) that will become apparent as a delay of several seconds between the highest frequency IF and the lowest. This would be measurable if the bursts had some sufficiently sharp feature.
An observation at 21 cm was performed to
  make use of the lower Galactic confusion and
  high sensitivity of the WSRT.
  We reanalyzed five archival VLA datasets taken between 1986 and 1989 and the 2002 discovery dataset. All of these except the latter were pointed at SgrA. Two of them, both obtained in A-configuration, had not been imaged before with the proper three-dimensional image restoration techniques. The complete set of observations we reduced is specified in table~\ref{table0}.
 
\section{Data reduction}
\subsection{General}
We used AIPS \citep{Greisen} for the reduction of all datasets.
\subsection{The 92 cm WSRT observations on 2005 March 24}

\begin{figure}
     \resizebox{\hsize}{!}{\includegraphics{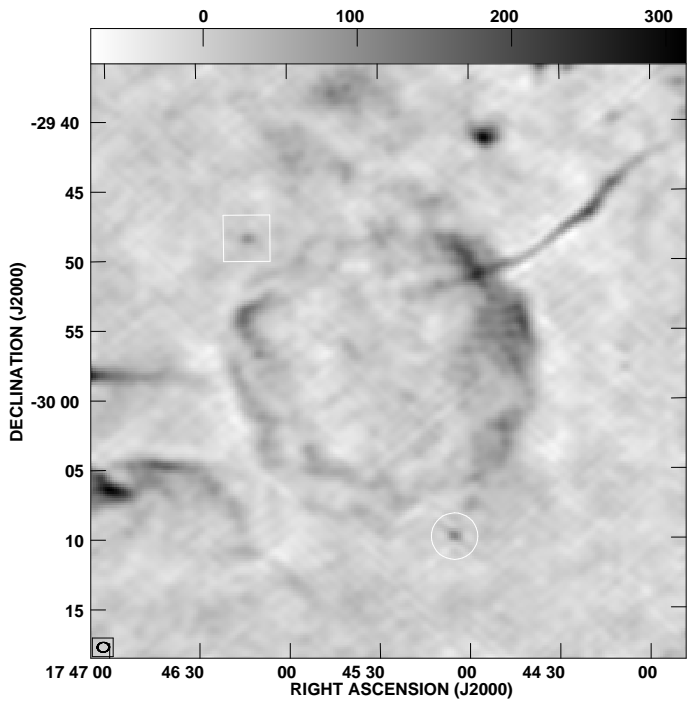}}
     \caption{The supernova remnant G359.1-0.5 with "The Snake" to the northwest, from our reduction of the GCRT J1745-3009 discovery observation on 2002 September 30/October 1 with the VLA in CnB configuration. This observation
              revealed this transient, indicated by a circle, for the first time \citep[see][]{Hyman1}. Noise levels in this image vary from 5 to 13 mJy $\mathrm{beam^{-1}}$ across the image. A Gaussian fit to the unresolved GCRT J1745-3009 gives a peak flux density of  only 116$\pm$14 mJy 
              $\mathrm{beam^{-1}}$ because the five Jy-level bursts have been averaged over about 6h of observation. A Gaussian fit to the source to the northeast 
              of the supernova remnant, 
              indicated by the box, gives a peak flux density of 91$\pm$14 mJy $\mathrm {beam^{-1}}$. Correction for primary beam attenuation has been applied.}
     \label{snremnant2002sep30}
\end{figure}

\begin{table*}
\caption{Specifications of these observations.}
\label{table0}
\centering
\begin{tabular}{c c c c c c c c c c c}
\hline \hline
No. & Date    & Telescope& Number  & Number& Number& Bandwidth   & Number of & Ch.width & Tot.BW& On-source\\
    & (yymmdd)& (+conf.) & of      & of    & of chann. & per IF & pol.prod. &  for ima- & for ima-& time (h) \\
    &         &          & antennas$^{\mathrm{1}}$& IFs   $^{\mathrm{2}}$& per IF   & (MHz)  & per IF& ging (kHz)& ging (MHz)$^{\mathrm{3}}$& \\
\hline
1   &  860329 &  VLA A   & 11       & 1     & 127       & 3.1         & 1    & 98 & 2.5 &4.6\\
    &        &          &          &       &           &             &      & & &\\
2   & 860805 &  VLA B   & 8        & 1     & 127        & 0.8      & 1      & 98& 0.7 & 4.9\\   
    &          &          &       &           &             &      & & &\\
3   & 861226 &  VLA C   & 15       & 1     & 63         & 0.8       & 1     & 98 & 0.7 & 6.2\\
    &        &          &           &       &            &             &    & & &\\
4   & 881203 &  VLA A   & 22        & 2     & 7         &  1.4       & 1    & 195 & 2.7 &5.7\\
    &        &         &           &                    &              &    &   & &\\
5   & 890318 & VLA B    & 27       & 2      & 7         &  0.7       & 1    & 98 & 1.4 & 5.3\\
    &        &          &          &        &           &              &    &  & &\\
6   & 020930 & VLA CnB  & 22       & 2      & 31       & 3.0       & 2      &  98 & 8.2 & 5.3\\
    &        &          &          &        &           &              &    &  & &\\
7   & 050324 & WSRT     & 12       & 7      & 128       & 10.0       & 4    &  6328 & 89& 5.3\\
    &        &          &          &        &           &              &    & & &\\
8   & 050514 & WSRT (21cm)& 14     & 6      & 64        & 20.0       & 4    & 12813& 154 & 4.6\\
    &        &           &         &        &           &              &    & & & \\
\hline
\end{tabular}
\begin{list}{}{}
\item[$^{\mathrm{1}}$] This is the nummer of antennas after flagging.
\item[$^{\mathrm{2}}$] This is the number of IFs after flagging averaged over the RR and LL polarization products, if both are available.
\item[$^{\mathrm{3}}$] This is total bandwidth for Stokes I imaging, we added RR and LL bandwidth.
\end{list}
\end{table*}

\begin{table*}
\caption{Flux measurements at 92~cm (unless otherwise noted) for detections and nondetections of GCRT J1745-3009 at $\mathrm{\alpha=17h45m5.15s, \delta=-30\degr09\arcmin52.7\arcsec}$ \citep{Kaplan}. Corrections for primary beam attenuation and bandwidth smearing have been applied where appropiate.}
\label{table2}
\centering
\begin{tabular}{c c c c c c c}
\hline \hline
No. & Date             & Telescope & Peak flux density  & error on fit & rms noise & resolution \\
    & (yymmdd)         & (+conf.)  & (mJy/beam)         & (mJy/beam)   & (mJy/beam)& ($\arcsec\times \arcsec$) \\
\hline
1   & 860329           &  VLA A    &   -49              & 27           & 18        &  $10\times 4$ \\
    &                  &           &                    &              &           &                             \\
2   & 860805           &  VLA B    & -19                & 20           & 100 $^{\mathrm{1}}$       & $44\times 33$  \\   
    &                  &           &                    &              &           &                              \\
3   & 861226           &  VLA C    & -26                & 29           & 100$^{\mathrm{1}}$        & $105\times 54$ \\
    &                  &           &                    &              &           &                          \\         
4   & 881203           &  VLA A    & -18                &  15          & 15        &  $12\times 6$ \\
    &                  &           &                    &              &           &                             \\
5   & 890318           & VLA B     & 41                 & 19           & 8         & $27\times 14$  \\
    &                  &           &                    &              &           &                   \\
6   & 020930           & VLA CnB   & 110 $^{\mathrm{2}}$& 8  $^{\mathrm{2}}$          & 8 $^{\mathrm{3}}$        & $44\times 36$ \\
    &                  &           &                    &              &           &                   \\
7   & 050324           & WSRT      & 5                  & 4            & 4         & $148\times 27$ \\
    &                  &           &                    &              &           &                   \\
8   & 050514           & WSRT (21cm)&-0.3               & 0.2          & 0.2       & $68\times 9$  \\
    &                  &           &                    &              &           &                   \\
\hline
\end{tabular}
\begin{list}{}{}
\item[$^{\mathrm{1}}$] The formal rms noise levels in these two maps are 19 mJy $\mathrm {beam^{-1}}$ and 69 mJy $\mathrm {beam^{-1}}$ for the 1986 August 5 and December 26 observations respectively, (much) lower than the indicated value of 100 mJy $\mathrm {beam^{-1}}$. However, many bright compact sources that should be detectable in these maps, are not due to the very poor uv coverage of this observation. We accounted for this by replacing the rms noise by a higher number, in this way giving a very crude representation of these missing sources. 
\item[$^{\mathrm{2}}$] Here we did not tie the clean beam fit to the position from \citet{Kaplan}, but set the AIPS task 'IMFIT' to solve for peak flux density as well as position in the residual image.
\item[$^{\mathrm{3}}$] This is the average noise in the residual image.
\end{list}
\end{table*}

   The WSRT 92 cm observations on 2005 March 24  started at UT 01:22 with the
   observation of the calibration source 3C295. We acquired data from the target field
   from 02:33 until 07:50 using 10s integrations, with eight 10-MHz IFs, consisting of 128 channels, each 78.125 kHz wide, separated 8.75 MHz from each other and
   centered on frequencies ranging from 315.4 to 376.6 MHz.  RFI was excised from the 
   spectral line data using the AIPS task 'SPFLG', while remaining RFI was removed from the
   continuum data using the AIPS task 'TVFLG'. \\
   Calibration was done in four steps. First we determined the variation in system
   temperature as a function of time (and therefore also position on the sky),
   using the intermittent firing of a stable noise source. Next we performed a bandpass calibration using the AIPS task 'BPASS'. We applied the bandpass solution using the AIPS task 'SPLAT', producing a continuum file with one channel per IF. After that, we performed
   an external absolute gain calibration using an assumed flux of 61.5 Jy
   for 3C295 in the lowest frequency IF, by running the AIPS tasks 'SETJY' and 'CALIB'. 'SETJY' was set to use the
   absolute flux density calibration determined by \citet{Baars} and the latest (epoch 1999.2) 
   polynomial coefficients for interpolating over frequency as determined at the VLA by NRAO staff.
   Finally, we self-calibrated the data for time variations in the
   relative complex gain phase and amplitude. \\
Theoretically, we should be able to reach a thermal noise
   level of 0.15 mJy/beam in a 5 hour integration, or at least the nominal
   beam confusion noise limit of 0.3 mJy/beam. However, we did not attain this
   sensitivity due to the limited uv-coverage, RFI, and the existence of
   bright diffuse emission in the field. The latter compromises both
   self-calibration and image quality. This could be remedied to some extent by
   excluding spacings below a certain limit ($\mathrm{uv_{min}}$ $>$ some multiple of $\lambda$, the wavelength). We chose a
   $\mathrm{uv_{min}}$ of $\mathrm {1.0 k\lambda}$ to eliminate the bulk of the diffuse emission, which could not be 
   deconvolved with the available uv-coverage.
   SgrA and Tornado are the dominant sources in the
   field, their sidelobes contributed
   significantly to the image noise level of 9.0 mJy $\mathrm {beam^{-1}}$ at the location of
   GCRT J1745-3009. These and other sources were deconvolved in an image with an asymmetrical cell size 
   ($10\arcsec\times 60\arcsec$). We chose to do so
   because a symmetrical cell size would yield a very elongated synthesized beam, this would hamper the deconvolution
   process. We subtracted the clean components of all sources from the uv-data before imaging the residual data with a symmetrical cell size. To lower the noise from the sidelobes of the two poorly subtracted extended sources,
   this final residual image was made by imposing a more severe lower limit of
   $2.5 \mathrm{k\lambda}$ on the spacings, which resulted in a noise level of
   3.7 mJy $\mathrm {beam^{-1}}$. That final image was made from only $7\%$ of the recorded visibilities.\\ 
   In retrospect, it is possible that the self-calibration process was adversely affected by bandwidth smearing, particularly because SgrA and Tornado were located far from the phase tracking center. Bandwidth smearing could have been diminished by keeping many channels per IF in 'SPLAT'. SgrA and Tornado were close to the half power beam width (HPBW). This also hampers self-calibration because the frequency dependence of the primary beam attenuation is much stronger near the HPBW than near the pointing center. It could have been fixed to some extent by running self-calibration per IF, at the expense of signal to noise.
   These flaws, the poor uv-coverage, the exclusion of many spacings and the
   Galactic plane contribution to the system temperature explains why the
   achieved noise level is still well above the thermal noise limit of 0.68 mJy $\mathrm {beam^{-1}}$ for this number of visibilities, 
   imaging bandwidth and IFs (see table~\ref{table0}), for a circular $60\arcsec$ beam towards cold sky. 
  
\subsection{The 21cm WSRT observations on 2005 May 14/15}
The 2005 May 14/15 observations at 21 cm started at UT 22:33 with the observation of the calibration source 3C286. We acquired data from the GCRT
   J1745-3009 field from 23:09 until 03:46 using 10 s integrations, with
   eight 20-MHz IFs, separated 17 MHz from each other and centered on frequencies ranging from 1265 to 1384 MHz. The
   calibration was done in the same way as for the 92 cm WSRT observation. The assumed flux for the calibrator source
   3C286 in the lowest frequency IF was 15.6 Jy.
   Theoretically, the rms sensitivity of these observations could be as low as
   about 21 $\mathrm{\mu Jy}$ $\mathrm {beam^{-1}}$, for a 4.6 h integration.  However, as for the 92 cm WSRT data,
   we excluded short spacings to eliminate most of the diffuse emission, which was necessary for successful
   self-calibration.
   The rms noise level in the final residual image
   was about 210 $\mathrm{\mu Jy}$ $\mathrm {beam^{-1}}$. That noise level is partly due to the loss of data: the exclusion of spacings below $2.5\mathrm{k\lambda}$ and the excision of RFI.
   The total loss of visibilities up to the final image was as high as $55\%$. With this number of visibilities
   and with the imaging bandwidth and IFs as mentioned in table~\ref{table0}, the theoretical thermal noise limit 
   is 45 $\mathrm{\mu Jy}$  $\mathrm {beam^{-1}}$ for a circular $13\arcsec$ beam towards cold sky.

\subsection{The 92 cm VLA discovery dataset of 2002 September 30/October 1}
The specifics of the 2002 discovery dataset are shown in table~\ref{table0}.
We started its reduction with the flagging of 4 of the 27 antennas. Also, we flagged individual spectral channels per baseline, per IF and per polarization product for all or part of the observing time, using the AIPS task 'SPFLG'.  We flagged small portions, of 1 minute or more, of data at the beginning and end of each scan using the AIPS task 'QUACK'. We also clipped data contaminated by RFI using the AIPS task 'CLIPM'. Next, we performed an external absolute gain calibration with an assumed flux of 25.9 Jy for 3C286 in the lowest frequency IF. This flux was determined by running the AIPS task 'SETJY', using the absolute flux density calibration determined by \citet{Baars} and the latest (epoch 1999.2) VLA polynomial coefficients for interpolating over frequency. We determined gain phase and gain amplitude solutions for both the primary calibrator 3C286 and the phase calibrator 1711-251, using the AIPS task 'CALIB'. This task was run using all spacings for the primary calibrator and spacings longer than $\mathrm{1k\lambda}$ for the phase calibrator. The AIPS task 'GETJY' determines the flux of the secondary calibrator from those gain solutions and the flux of the primary calibrator. 'GETJY' found a flux of 11.1 Jy for 1711-251 at the highest frequency IF (327.5 MHz). The gain solutions were interpolated using the AIPS task 'CLCAL'.\\
Next, we used 3C286 to find a bandpass solution. In doing so, we applied the interpolated gain solutions from 'CLCAL' for spacings longer than 500 wavelengths ($\mathrm{uv_{min}>0.5 k\lambda}$). For one of the antennas no visibilities were recorded during the scan of 3C286. Hence, no bandpass solution could be found for this antenna and only 22 antennas were left for imaging.
We applied the gain and bandpass solution to 20 of the total of 31 available channels using the AIPS task 'SPLAT'. Every two channels were averaged.\\
Next, we performed 18 iterations of phase only self-calibration, using initial solution intervals of 5 minutes, gradually decreasing down to 1 minute. We used 195 kHz channels for imaging and a cellsize of $4\arcsec$. We used 85 $512\times512$ pixel facets to cover the primary beam and no facets for outlier fields. We performed an amplitude and phase self-calibration and we produced the final model from the spectral averaged dataset. After that, we reran 'SPLAT' on the line data, but this time without spectral averaging, selecting $21\times97$ kHz of the available channels. We phase self-calibrated the new dataset using the acquired model from the spectral averaged data. Next, we imaged and deconvolved our phase self-calibrated dataset using 61 facets to cover the primary beam and 22 facets for the outlier fields. This time we used $256\times256$ pixel facets with a pixel size of $10\arcsec$. We self-calibrated again, but this time we solved for amplitude and phase, using a solution interval of 1 minute. The total average gain was normalized in this process. We imaged and deconvolved 450 Jy of total flux from the amplitude and phase self-calibrated dataset to make our final model. Figure \ref{snremnant2002sep30} shows the central facet of this model after correction for primary beam attenuation. We noticed that SgrA is by far the brightest source in the field and that it is near the half power beam point. We anticipated that the calibration of the uv data could be optimized by applying separate gain solutions to the clean components of the facet with SgrA, so we ran the AIPS runfile PEELR on the clean components of the facet of SgrA, solving for gain amplitudes and phases on a timescale of 10s. We subtracted the clean components from the peeled data using the AIPS task 'UVSUB' and we determined the position of GCRT J1745-3009 in our final model using the AIPS task 'IMFIT'. We shifted the phase stopping centre to this position using the AIPS task 'UVFIX' and we averaged all spectral channels using the AIPS task 'SPLIT'. We did a final edit using the AIPS task 'CLIP' and set $\mathrm{uv_{min}=1.0 k\lambda}$. We ran the AIPS task 'DFTPL' on this final residual dataset to produce our lightcurves. We did not correct the output of 'DFTPL' for primary beam attenuation because GCRT J1745-3009 was about $13\arcmin$ from the pointing center. Primary beam attenuation for this angular separation is only $1.8\%$.\\
In retrospect, it turned out that both the amplitude and phase (A\&P) self-calibration and the peeling of SgrA had negligible effect on the burst shapes in the final lightcurves. So the dataset could be reduced in a standard way, except perhaps for the large number of selfcal iterations and the exclusion of a rather large number of antennas, 5 of the 27 antennas being excluded for the entire observation. 

\section{The source on the opposite side of the supernova remnant}

The source northeast of the supernova remnant G359.1-0.5, indicated by a box in figure~\ref{snremnant2002sep30} is resolved in VLA A configuration. From a combination of three VLA datasets, two in A configuration and one in B configuration, this source was detected with a peak flux density of 17.1$\pm$2 mJy $\mathrm {beam^{-1}}$ and an integrated flux of 47.6 mJy \citep[see][ table 2, source 72]{Nord2004}.  Apparently the synthesized beam of the combination of these datasets ($12\arcsec\times 7\arcsec$) resolves this source. As noted in the caption of figure~\ref{snremnant2002sep30}, the peak flux density we derived from the 2002 discovery observation is 91$\pm$14 mJy $\mathrm {beam^{-1}}$. A large fraction of the difference with the integrated flux measurement by \citet{Nord2004} is probably caused by extended emission. Indeed, when we exclude the shortest spacings, $\mathrm{uv_{min}}=\mathrm {1.0 k\lambda}$, we find a much lower peak flux density of 73$\pm$5 mJy $\mathrm {beam^{-1}}$. The remaining difference may also come from extended emission that is picked up differently by these observations. \\
However, the main reason that this source drew our attention is its absence in a high dynamic range image of the Galactic Centre at 92 cm with a noise level of about 5 mJy $\mathrm {beam^{-1}}$ and an angular resolution of $43\arcsec$ \citep[see][ figure 11, hereafter called the LaRosa map]{LaRosa}. The datasets used for the LaRosa map were taken on 1986 August 5 (B conf., 8 antennas) and 1986 December 26 (C conf., 15 antennas), 1987 March 25  (D conf., 15 antennas) and 1989 March 18 (B conf., 27 antennas).  Our reduction of the 1989 March 18 data shows the source at the $\geq 6\sigma$ level, a Gaussian fit gave a peak flux density of 53$\pm$7 mJy $\mathrm {beam^{-1}}$. Here, the size of the synthesized beam is $27\arcsec\times 14\arcsec$ while we set $\mathrm{uv_{min}}$ to $\mathrm {2.0 k\lambda}$. This clear detection indicates that the non-detection of the source in the LaRosa map is probably not due to transience. More likely, the source is concealed in the LaRosa map by a negative background peak.\\

\begin{figure*} 
\centering
     \includegraphics[width=17cm]{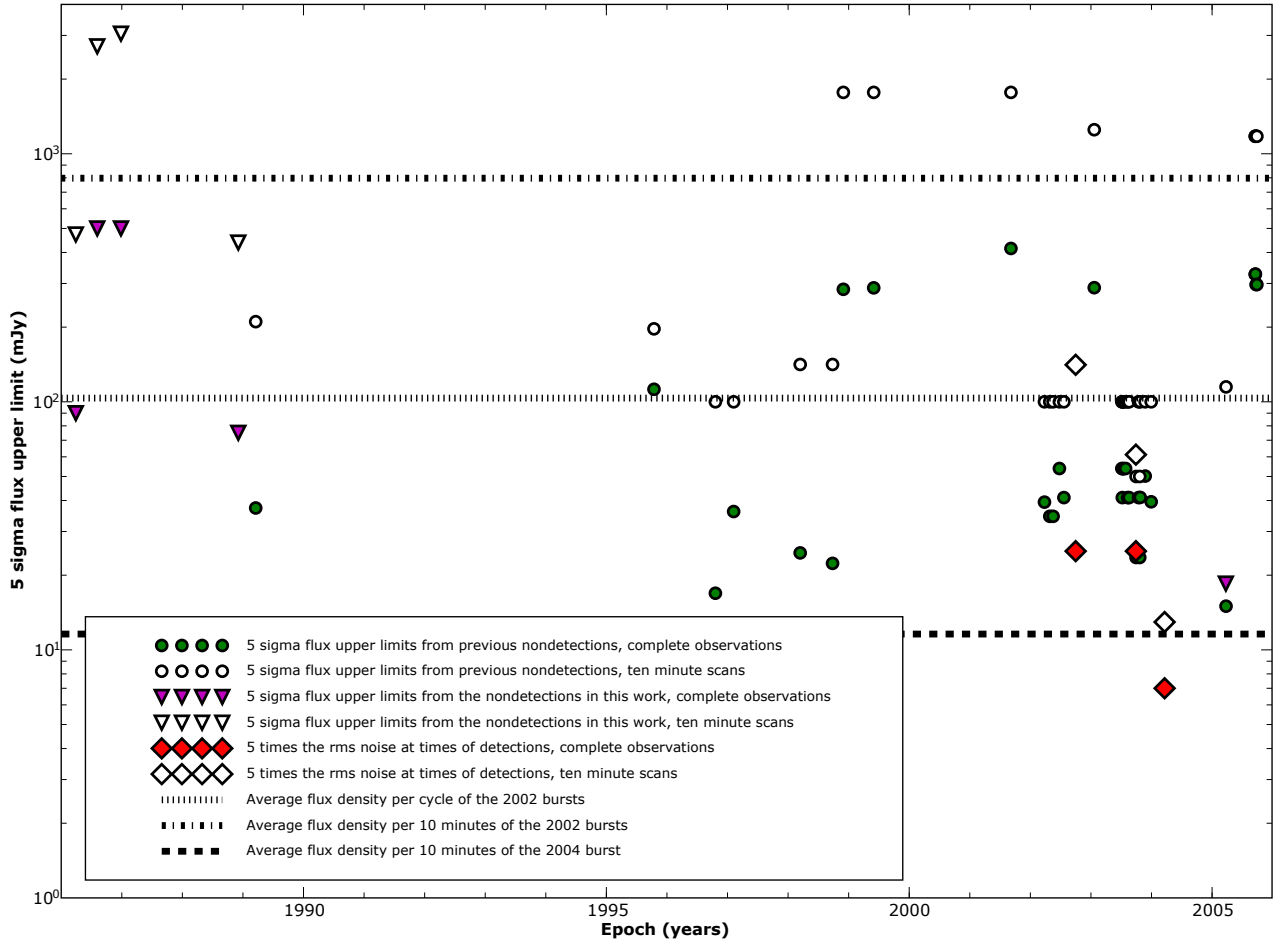}
     \caption{Approximate detection thresholds ($5 \sigma$ noise levels)
              at the location of GCRT J1745-3009 
              of 41 Galactic Center observations at 92 cm 
              over two decades.   
              For the WSRT observation at 92 cm, the 10 minute scan sensitivity is not indicated, 
              since the snapshot point spread function (psf) of a linear array does 
               not allow to do this accurately.
              The observations in this plot
              start on 1986 March 29 and end on 2005 September 27. 
              }
     \label{noise_comparison}
\end{figure*}

\section{Overview of flux measurements of GCRT J1745-3009}
We hoped to redetect GCRT J1745-3009 with the WSRT, with some of the VLA observations mentioned in the previous section and with two additional A configuration observations from the VLA archive. We did not redetect the source, but we measured its flux at its position in all of the seven maps. Specifics of these observations are shown in table~\ref{table0}. Note that the on-source time for the two WSRT observations is comparable to the VLA observations, despite the limited time for which the WSRT can observe this low declination source. The reason for this is that the WSRT in general does not need to observe secondary calibrators. The results of the flux measurements at these epochs and at the time of the discovery are shown in table~\ref{table2}. For the seven nondetections, we fitted the restoring beam to the position reported by \citet{Kaplan}. We have also imaged 10 minute subsets of the residual data from the five 1986-1989 observations to look for isolated bursts, but we found none. \\
We merged our results from table~\ref{table2} with those from a recent overview of observations since 1989 \citep[see][table 1]{Hyman2} together with the results from the second redetection \citep{Hyman3} to produce a plot of $5\sigma$ flux upper limits on quiescent emission from GCRT J1745-3009 in the 92 cm band (see fig.~\ref{noise_comparison}). In order to derive appropriate values, we scaled the 10-minute scan sensitivities mentioned (20 and 10 mJy $\mathrm {beam^{-1}}$ for the VLA and the GMRT respectively, after correction for primary beam attenuation) with the square root of the observing bandwidth, taking 6.2 MHz as the base. The sensitivities for complete observations were also scaled with the square root of the total on-source time. We note that the 1989 March 18 observation was already analysed by \cite{Hyman2} and their reduction led to slightly more constraining values, so we adopted these in figure~\ref{noise_comparison}. Here, we took account of the fact that the total bandwidth of that observation was actually 1.4 MHz instead of the 12.5 MHz mentioned in their table 1. Consequently, we derived $5\sigma$ upper limits of $5\cdot20\cdot\sqrt{6.2/1.4}=210$ mJy $\mathrm {beam^{-1}}$ and  $5\cdot20\cdot\sqrt{6.2/1.4}/\sqrt{5.3\cdot 6}=37$ mJy $\mathrm {beam^{-1}}$, for those 10-minute scans and for that complete observation, respectively. \\
The lowest noise level of all 92 cm observations, about 6 mJy $\mathrm {beam^{-1}}$ in a 2 minute interval, was achieved at the time of the second redetection, with the GMRT on 2004 March 20 \citep{Hyman3}. This is actually the only observation that could have detected bursts of this kind and only by making 2 minute scan averages. None of the observations included in figure \ref{noise_comparison} can detect the 2004 burst \citep{Hyman3} in 10 minute averages at the $5 \sigma$ noise level.\\
The WSRT 2005 May 14/15 5$\sigma$ upper limit at 21cm (1.05 mJy $\mathrm {beam^{-1}}$) was less constraining than the VLA upper limit at that wavelength on 2005 March 25 \citep[0.4 mJy $\mathrm {beam^{-1}}$, see][]{Hyman2}. 21 cm observations are not included in figure~\ref{noise_comparison}.

\section{Reanalysis of the 2002 discovery dataset}

\begin{figure*} 
\centering
     \includegraphics[width=17cm]{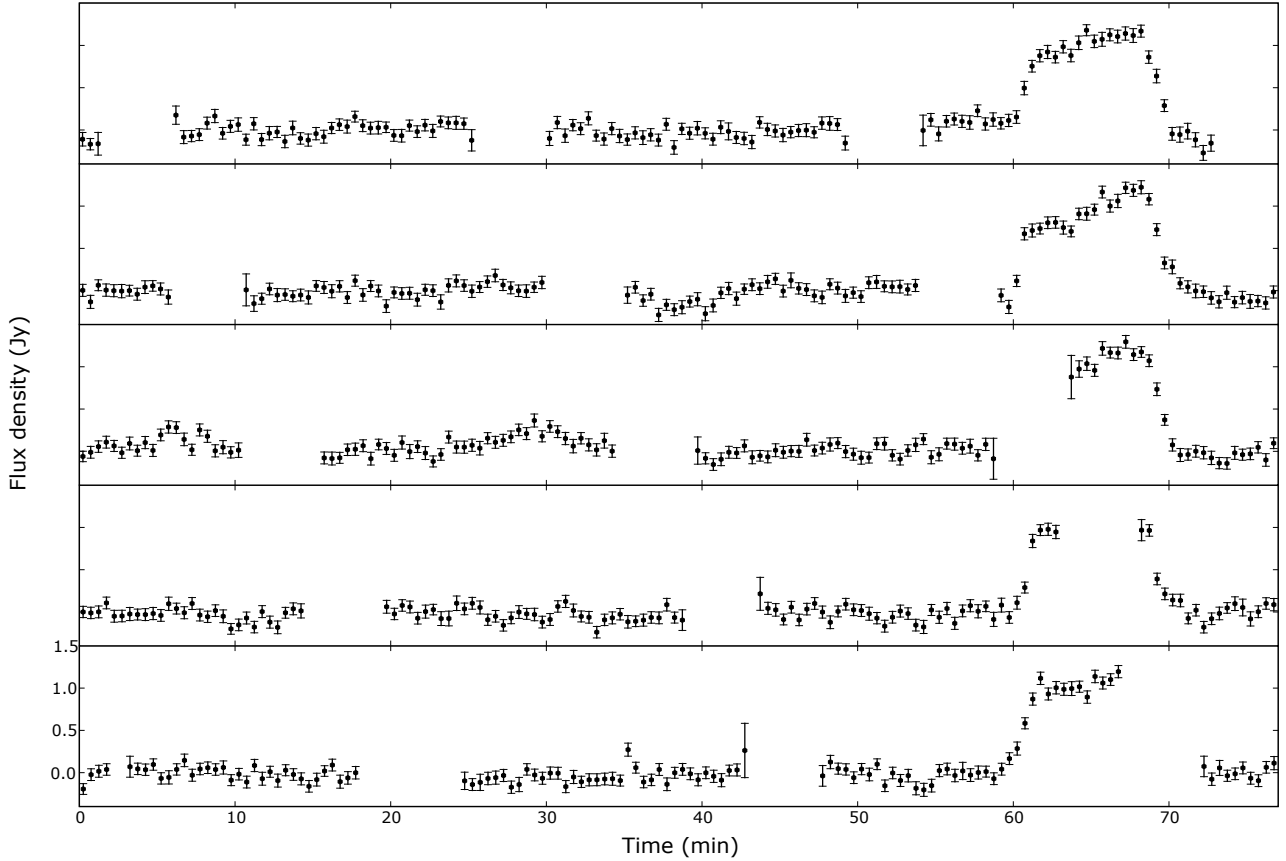}
     \caption{The plot above shows the lightcurve from the discovery dataset of GCRT J1745-3009 with 30s sampling. This plot is setup in the same way as the lightcurve in the discovery paper except for the flux density measurements between bursts. For those nondetections \citet{Hyman1} showed $3\sigma$ upper limits on interburst emission, we show the actual background flux density measurements. Also, we have folded the lightcurve at intervals of 77.012 minutes instead of 77.130 minutes. The first interval is shown in the bottom panel, starting at 20h50m00s on 2002 September 30 (IAT). The average of all the error bars shown is 74 mJy. The gaps are due to phase calibrator observations.}
     \label{lightcurves}
\end{figure*}

\begin{figure*} 
\centering
     \includegraphics[width=17cm]{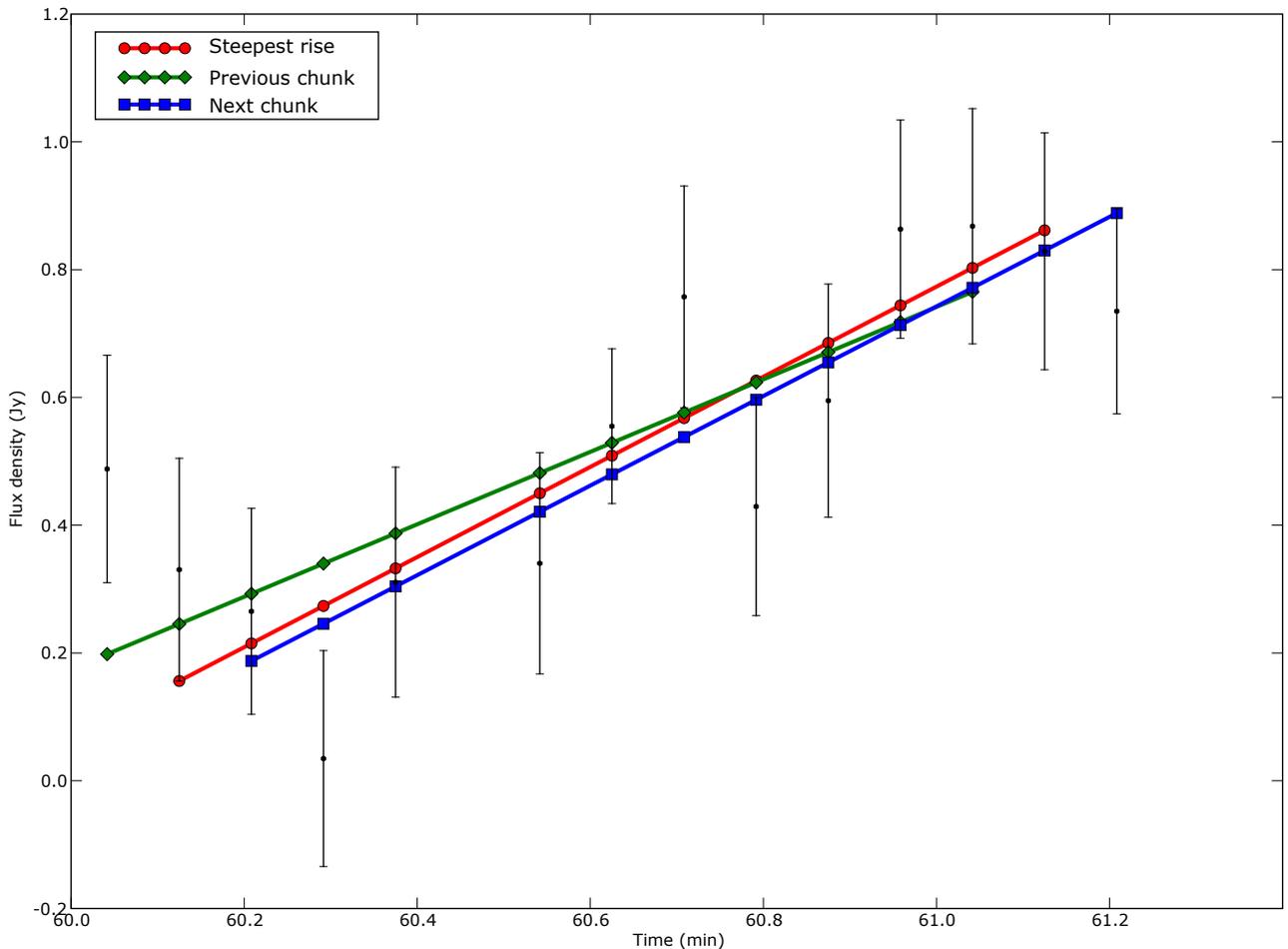}
     \caption{This plot illustrates how the times of steepest rise for four of the bursts are determined. Weighted linear regression is performed on successive one minute chunks of data. The chunks have a maximum of 55s of overlap time. Here the rising part of the first burst is shown.}
     \label{fig:steepestrise}
\end{figure*}

\begin{figure*} 
\centering
     \includegraphics[width=17cm]{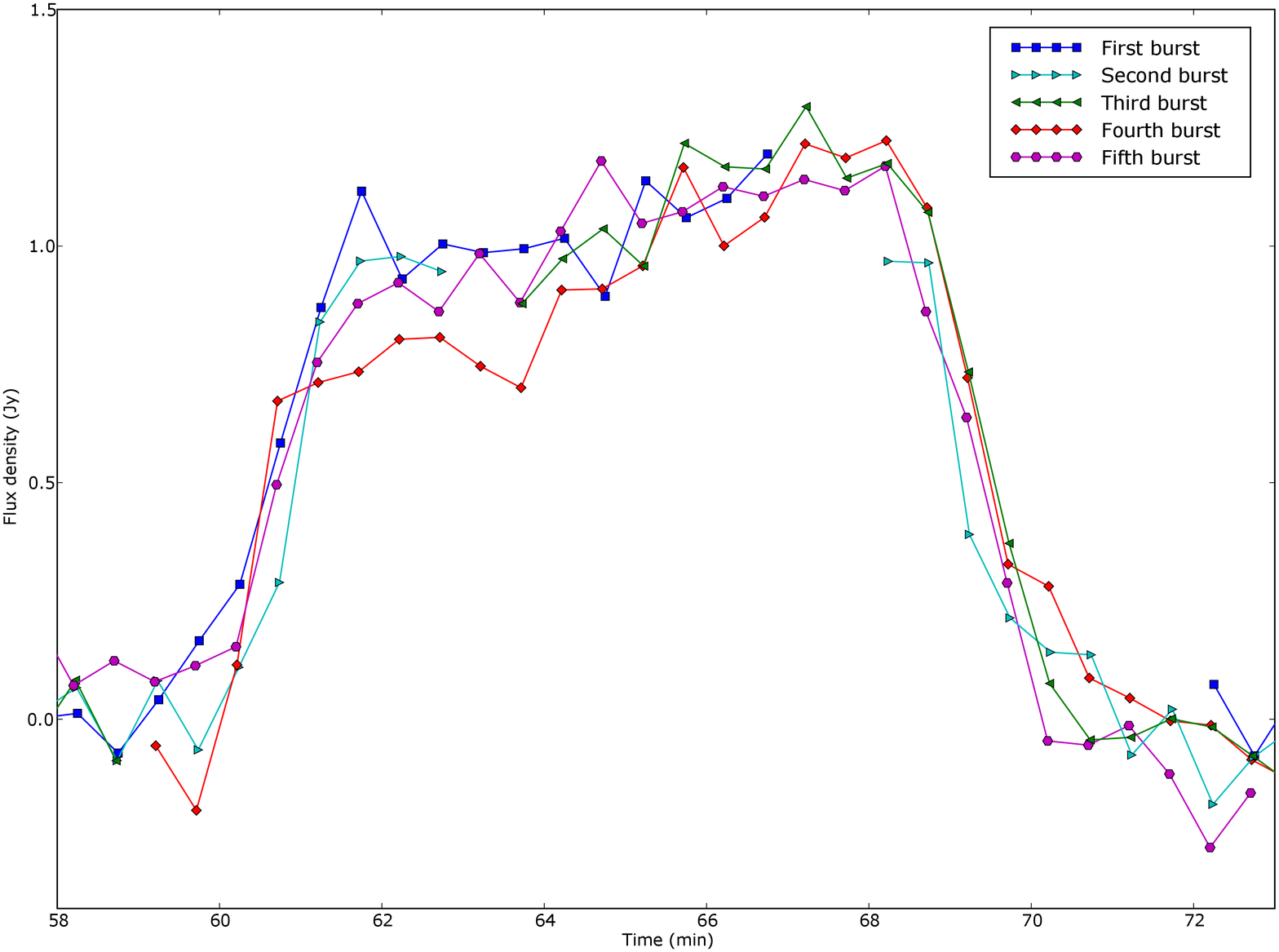}
     \caption{The plot above shows the five bursts from the discovery dataset of GCRT J1745-3009 with 30s sampling folded at intervals of 77.012 minutes. Time is relative to 20h50m00s on 2002 September 30 (IAT) (plus multiples of 77.012 minutes).}
     \label{allburstsoneplot}
\end{figure*}

\subsection{Lightcurve}
The lightcurve that we extracted from the discovery dataset of GCRT J1745-3009 at the position derived in paragraph \ref{subsec:pos} is shown in figure~\ref{lightcurves}. The bursts seem to have similar shapes: a steep rise, a gradual brightening and a steep decay, more consistent than the bursts shown in figure 1 of the discovery paper \citep{Hyman1}. This lightcurve is twice as accurate as the original one. We also ran the AIPS task 'DFTPL' with 5s sampling, this is the integration time for the recording of the visibilities in the discovery dataset. We found no compelling evidence for interburst emission, not even on the shortest (5s) timescale. We determined the recurrence interval between bursts by measuring the times of steepest rise for four of the bursts. Consecutive 1 minute chunks of data were selected by a sliding window. For each chunk of data we determined its average slope by weighted linear regression. The weights come from the reciprocal of the noise variances from 'DFTPL'.  The time corresponding to the steepest positive slope was then calculated as the weighted average of the timestamps in the datachunk. For the first burst, this method is illustrated  in figure \ref{fig:steepestrise}. Weighted linear regression also calculates the error bars of the times of steepest rise from the error bars of the data points. The times of steepest rise and the corresponding error bars are shown in table~\ref{table3}. The times mentioned in that table are relative to 20h50m00s on 2002 September 30 (IAT). We then again applied the formulae for weighted linear regression to find the period between bursts and its $1\sigma$ error. We found a period of 77.012$\pm$0.021 min from the values in table~\ref{table3}. We have improved the error on the period by an order of magnitude \citep[][paragraph 3 and caption of figure 3]{Hyman2}, but the period itself agrees with the previously determined period of 77.1 m $\mathrm{\pm15s}$. However, it is important to note that our method differs from the one used by \citet{Hyman1}. We have made no assumption with regard to the burst shapes in determining the period. \\
The residuals with respect to that fit are 0.097, -0.114, 0.053 and -0.007 minutes for the first, second, fourth and fifth burst, respectively. The residual for the second burst is the largest, 6.8s "too late" with respect to the fit, this corresponds to $1.9\sigma$, $\sigma=0.060$ min, this is the error on the time of steepest rise of the second burst. \\
We were also able to measure the times of steepest decay for four of the bursts in a similar manner, see table~\ref{table4}. For three bursts we could measure both the time of steepest decay and the time of steepest rise. In this way we found that the time between steepest rise and steepest decay varies. We found intervals of 8.29$\pm$0.08, 8.87$\pm$0.09 and 8.66$\pm$0.09 min for the second, fourth and fifth burst, respectively. So for the second burst the interval between steepest rise and steepest decay is 3.45$\%$ less than the weighted mean of those three intervals. The significance of this deviation is $3.0\sigma$.\\
We can use the derived period to fold the bursts in one plot, see figure~\ref{allburstsoneplot}. This plot shows that the bursts indeed have similar shapes.\\
\begin{table}
\caption{Measurements of times of steepest rise for four bursts}
\label{table3}
\centering
\begin{tabular}{c c c c c}
\hline \hline
Burst & Time of steepest& 1$\sigma$ error & Slope   & 1$\sigma$ error \\
number & rise (min)     & (min)           & (Jy/min)& (Jy/min)\\
\hline
1            & 60.624                &  0.068          &   0.706 & 0.158\\
2           & 137.848                &  0.060          &   0.828 & 0.175\\   
4           & 291.704                &  0.065          &   0.724 & 0.138\\      
5           & 368.776                &  0.065          &   0.743 & 0.150\\
\hline
\end{tabular}
\end{table}

\begin{table}
\caption{Measurements of times of steepest decay for four bursts}
\label{table4}
\centering
\begin{tabular}{c c c c c}
\hline \hline
Burst & Time of steepest & 1$\sigma$ error & Slope  & 1$\sigma$ error\\
number & decay (min)                 & (min)           & (Jy/min) & (Jy/min) \\
\hline
2            & 146.136                &  0.041          &   -1.125 & 0.146 \\
3           & 223.489                &  0.057          &   -0.811 &  0.150\\   
4           & 300.578                &  0.067          &   -0.717 & 0.150\\      
5           & 377.439                &  0.066          &   -0.734 & 0.148 \\
\hline
\end{tabular}
\end{table}

\subsection{Implications for other observations}
Now that we have determined the periodicity of the bursts more accurately, we can check if other short GC observations at 92 cm before and after the discovery observation should have detected GCRT J1745-3009. The observation closest in time was taken on 2002 July 21 \citep[see][table 1]{Hyman2}. This was a 59.2 minute scan starting 1719.75 hours before the start of the bright part of the first burst in the discovery dataset. This corresponds to 1339.86 periods of 77.012 minutes. Consequently, the source should not have been seen during that short scan and this was indeed the case \citep{Hyman2}. However, there is a large uncertainty in calculating burst times over an interval as large as 71 days. The error is $\mathrm{0.021min\cdot1339=28}$ min. From that uncertainty and Gaussian statistics, we calculated that the chance of having observed at least 5 minutes of bursting activity on 2002 July 21 was $74\%$, assuming that GCRT J1745-3009 were bursting as during the discovery observation. If GCRT J1745-3009 was indeed active on 2002 July 21, we can infer from the nondetection on that occasion that P, the recurrence interval between bursts is tightly constrained: $\mathrm{77.007 min<P<77.021min}$. \\
The next observation closest in time was taken on 2002 June 24. Its duration was only 34.5 min, starting 1842.17 periods of 77.012 minutes before the start of the bright part of the first burst in the discovery dataset. During this observation we should have seen at least 6 minutes of a burst if we take into account the constraints on the period from the nondetection on 2002 July 21. From the fact that we did not detect emission on 2002 June 24 we may conclude that activity started after this 34.5 minute scan. \\
The first suitably pointed 92 cm observation after the discovery observation was taken on 2003 January 20. The source was not detected, but the data were taken with the VLA in CD configuration. This implies that rms noise levels from 10-min scans are about 250 mJy $\mathrm {beam^{-1}}$ \citep[see][and figure~\ref{noise_comparison} in this paper]{Hyman2}. Thus it is likely that GCRT J1745-3009 could not have been detected at the $5\sigma$ level on 2003 January 20, even if an individual ten minute scan were spaced in time such that it completely covered a burst. It may be that the activity continued until the summer of 2003 when three 59 minute and four 34 minute GC observations were performed with the VLA in A configuration. At least two of these scans are spaced in time such that if one covered the interval between two bursts, the other must have covered a complete burst. So we are sure that the recurrent bursting activity of GCRT J1745-3009 stopped before it was redetected on 2003 September 28.\\
In summary, the bursting activity with a period of 77.012 minutes as seen during the discovery observation must have started after 2002 June 24 and may have continued until the summer of 2003. Unfortunately, we cannot constrain the timespan of a recurrently bursting GCRT J1745-3009 to less than a year.

\subsection{Position and flux measurements; spectral index determination}
\label{subsec:pos}
The most accurate position measurement, corresponding to the highest signal to noise ratio, can be achieved by selecting just the time intervals that cover the bursts.  We found a peak flux density of $900\pm23$ mJy $\mathrm {beam^{-1}}$ and this J2000 position: $\mathrm{\alpha=17h45m05.015s\pm0.045s, \delta=-30\degr09\arcmin52.19\arcsec\pm0.52\arcsec}$. This position of GCRT J1745-3009 has not yet been corrected for ionospheric-induced refraction \citep[see][for some background]{Nord2004}. That correction, which is basically, but not exactly, a global position shift of all sources in the field, will significantly increase the uncertainty in the position of GCRT J1745-3009. Here, we just mention that in our maps the bright source SGR E46 is 0.33s west and 0.89$\arcsec$ north of the NVSS \citep{NVSS} position. The NVSS catalogue mentions a positional accuracy of $0.45\arcsec$ in right ascension and $0.6\arcsec$ in declination for this source. We consider the actual uncertainty for the given position of GCRT J1745-3009 to be $5\arcsec$ in both right ascension and declination.\\
Rms noise values in the map that constitutes our final model range between 5 and 13 mJy $\mathrm {beam^{-1}}$. We also made a map from the same data, but without short spacings ($uv_{\mathrm {min}}=\mathrm {1.0 k\lambda}$). Noise levels then drop significantly, varying between 4 and 6 mJy $\mathrm {beam^{-1}}$ across the image. We removed the bursts and we made a cleaned image with the same spacings. The noise levels are somewhat higher now: between 5 and 7 mJy $\mathrm {beam^{-1}}$.\\
In order to derive an upper limit on interburst emission we fitted the clean beam to the position measured above. We found a peak flux density of $-0.6\pm6.4$ mJy $\mathrm {beam^{-1}}$, after correction for primary beam attenuation ($1.8\%$). This gives a $5\sigma$ upper limit on interburst emission of 31 mJy $\mathrm {beam^{-1}}$. This is more than twice as constraining as the original upper limit.\\
Neglecting primary beam attenuation, we found a weighted mean flux of $103.5\pm2.9$ mJy $\mathrm{beam^{-1}}$ from the output of the AIPS task 'DFTPL' on the residual data with full (5s) sampling. We also ran 'DFTPL' on this data for each of the five bursts and for each of the two IFs separately. We only selected times for which both IFs had fluxes and then calculated the natural logarithm of the ratio of the fluxes for each timestamp and the variance of that quantity. We then calculated the weighted mean of these logarithms for each burst. The spectral index and error bar for each burst are shown in table~\ref{spectral_indices}, using the average frequencies of IF1 (327.5000 MHz) and IF2 (321.5625 MHz). The spectral indices and error bars of the individual bursts do not seem inconsistent with Gaussian statistics, so we calculated the weighted mean spectral index as well: $\alpha=-6.5\pm3.4$. This is not incompatible with the spectral indices found by \citet[][$\alpha=-4\pm5$ and $\alpha=-13.5\pm3.0$]{Hyman2,Hyman3}, given the large error bars. The weighted mean of these three measurements is $\alpha=-9.4\pm2.1$. 
\begin{table}
\caption{Measurement of spectral index for each burst}
\label{spectral_indices}
\centering
\begin{tabular}{c c c}
\hline \hline
Burst & $\alpha$ & 1$\sigma$ error \\
number & ($S_{\nu}\propto{\nu}^{\alpha}$) & \\
\hline
1           & -9.9                 & 6.7\\
2           & -9.0                 & 9.3 \\
3           & 0.9            &  8.7 \\   
4           & -0.4             &  6.9\\      
5           &    -12.3             &  6.9 \\
\hline
\end{tabular}
\end{table}

\subsection{Circular polarization}
We compared the lightcurves for left ('LL') and right ('RR') circular polarization with 30s sampling. Although there are occasional 'LL' and 'RR' flux differences during the bursts larger than the sums of the respective error bars, this is also seen in between the bursts. There is no compelling evidence for circularly polarized emission during any particular phase of the burst cycle. On the other hand, we cannot exclude it completely, because we have insufficient signal to noise in Stokes V. \\
From the residual data, we selected the times corresponding to the bursts and we made a Stokes V dirty image. We corrected for primary beam attenuation and fitted the clean beam to the position of GCRT J1745-3009 as we did in the previous paragraph to determine the upper limit for interburst emission . We measured a Stokes V of $-20\pm10$ mJy $\mathrm {beam^{-1}}$. Using the total intensity averaged over the bursts, $900\pm23$ mJy $\mathrm {beam^{-1}}$, we found that the $5\sigma$ upper limit on the fractional circular polarization, $\mathrm{|V|/I}$, is $8\%$. \citet{Hyman1} derived a weaker constraint of $15\%$ on the fractional circular polarization averaged over the bursts.\\
Despite the lack of evidence for circularly polarized emission in the discovery observation, it has been detected in the data from the 2003 recovery observation \citep{Roy2008}. Here, only the last part a single burst was covered. From this detection and the fact that the average of Stokes V over a complete burst (almost completely) vanishes we infer that during an earlier part of the burst, Stokes V must have the opposite sign. In other words, if we can assume that the 2003 burst is similar to the 2002 bursts with regard to circularly polarized emission, there must be a sign change in the circular polarization during the bursts.

\subsection{Maximum source size and maximum distance for incoherent emission}

All of the steep rising part of the bursts can be well approximated by a straight line. This is true even at the very beginning of the bursts, when the flux is at or just above the noise level. It can be seen in the lightcurve down to 10s sampling, but at full (5s) sampling we have insufficient signal to noise to trace any possible slope flattening down to the first 5s of the beginning of the bursts. The average slope of the bursts in table~\ref{table3} is 0.75 Jy/min or 0.125 Jy/10s. This implies a flux doubling time of $\mathrm{\Delta t=10s}$ at the beginning of the bursts, when the flux is 125 mJy. The maximum source size at that time is then 10 lightseconds, if we assume that the source is not moving at a relativistic velocity \citep[see, e.g.,][for some background]{Harris}. We can use the maximum source size $\mathrm c\cdot\Delta t$ to link the brightness temperature $\mathrm{T_b(K)}$ to the flux F and maximum distance D \citep[see, e.g,][]{Rybicki}:
\begin{equation}
\mathrm{T_b=\frac{\lambda^2 I_{\nu}}{2 k}=\frac{\lambda^2 F}{2 k \pi  \theta^2}=\frac{2 F}{\pi k} (\frac{D}{\nu \Delta t})^2}
\end{equation}
where $\mathrm{\lambda,I_{\nu},\nu}$, k and $\mathrm{\theta}$ are the wavelength, the specific intensity, the frequency, Boltzmann's constant and the angle subtended by the radius of the source, respectively. If we express the distance in pc, the flux in Jy and the frequency in GHz, we get:
\begin{equation}
\mathrm{T_b=4.39\cdot 10^{11} F  (\frac{D}{\nu \Delta t})^2}
\end{equation}
If synchrotron self-Compton radiation limits the brightness temperature to $\mathrm{{10}^{12}K}$, the maximum distance for a source of size ten lightseconds and a flux of 0.125 Jy emitting incoherently at 325 MHz is 14pc, assuming it is not moving at a relativistic velocity.
\citet{Hyman1} used the decay time of the bursts (conservatively estimated at $\simeq$ 2 min.) to calculate a maximum distance of 70 pc. So we have improved this upper limit by a factor 5.

\section{Discussion}
Five of these upper limits on the flux of GCRT J1745-3009 come from the oldest observations of this field in the 92 cm band. This may provide interesting constraints on the feasibility of the double neutron star binary model \citep{Turolla} in the near future. In this model, similar to J0737-3039, the period of recurrence of the 2002 bursts is explained by an orbital period of 77 minutes. The lack of activity for many years is explained by geodetic
precession, which could have caused the wind beam of the most luminous pulsar not to intercept the magnetosphere of the other pulsar for decades. \citet{ZhuXu} claim that the redetection in 2003 \citep{Hyman2} does not support this model. Their remark was, however, erroneously based on a geodetic precession period of $\simeq3$yr, but this is actually $\simeq21$ years\footnote{The $"$characteristic time for changing the system geometry$"$ as mentioned by Turolla et al. (2005) differs from the period of geodetic precession by a factor $2\pi$.}. The last redetection (2004 March 20) and the first observation (1986 March 29) are 18 years apart. Unfortunately this timespan is too short to test the double neutron star binary model, but not if we redetect the system in the near future. More constraining are the results from population synthesis models \citep[see, e.g.,][ fig.2]{Portegies}: fairly eccentric (0.3$<$e$<$0.6) double neutron star binaries with an orbital period of 77 minutes are scarce, even compared to systems like J0737-3039. Also, the unpulsed emission needed for this model has not been detected in J0737-3039 \citep{Chatterjee}.\\
The lightcurve from our reduction of the 2002 discovery dataset shows that the bursts have similar shapes. There are three distinct parts separated by breaks, a steep rise, a gradual brightening and a steep decay. The main differences with the lightcurve from the discovery paper \citep{Hyman1} can probably be explained by sidelobes from SgrA \citep[][end of section 2]{Roy}. These sidelobes are not seen in our images. Apparently, the lightcurve from the discovery paper was made by compiling fluxes from successive snapshot images \citep[][paragraph 2]{Hyman2}. Therefore, we also made a lightcurve with 30s sampling of the fourth burst using the AIPS task 'IMAGR' and natural weighting, but the differences were negligible. We also learned that the output from 'DFTPL' is likely to be more accurate than fluxes from snapshots (Eric Greisen 2009, private communication).\\ 
Our refined reduction of the discovery data seems to support
the transient white dwarf model pulsar proposed by \citet{ZhangGil}.
A light-house beam associated with a highly magnetized white dwarf can
emit radio emission with a 77 minute period while maintaining an
accuracy better than one second. The
duty cycle $9/77\simeq 0.1$ (with a few percent jitter from one pulse to another)
is typical for pulsars. Moreover, an intensity asymmetry between the opposite sides of single
pulses is typical in normal pulsars, so it can be expected also in
white dwarf pulsars. \\
On the other hand, if the bursts we see are actually convolved with some scattering function, the intrinsic shape of the bursts could be different. Interstellar scattering can cause bursts to decay exponentially. We compared exponential fits to weighted linear regression for the 1 min data chunks that we used to determine the times of steepest decay. We found that residuals for linear fits are slightly smaller (12$\%$ overall) than for exponential fits. The exponential fit was better than the linear fit for the tail of one of the four bursts only. From the exponential fits we found decay times of 0.56, 0.77, 0.73 and 0.81 min for the second, third, fourth and fifth burst, respectively. These values are rather large for a source near the GC. For an observing frequency of 325 MHz and for the position of GCRT J1745-3009 on the sky, pulse broadening times of 3.96-8.72s and a DM of 567-751$\mathrm{ cm^{-3}pc}$ are estimated from the NE2001 model of \citet{Cordes}, assuming a distance (to the GC) of 8 kpc \citep{Reid}. We also checked what dispersion measure would follow from our average scattering timescale (0.72 min) and the empirical relation found by \citet{Mitra}:
\begin{equation}
\mathrm{\tau_{sc}=4.5\cdot10^{-5}\cdot DM^{1.6}\cdot(1+3.1\cdot10^{-5}\cdot DM^3)\cdot \lambda^{4.4}}
\end{equation}
with the scattering time ($\mathrm{\tau_{sc}}$) in ms, the dispersion measure (DM) in $\mathrm{cm^{-3}}$pc and the observation wavelength ($\mathrm{\lambda}$) in meters.
From this relation we find a dispersion measure of $\simeq925$ $\mathrm{ cm^{-3}pc}$.  This would imply that GCRT J1745-3009 is located beyond the GC. For a check on consistency we compared this dispersion measure with the DM that can be found from the formula for the dispersion delay $\Delta t$ (in seconds):
\begin{equation}
\mathrm{\Delta t=4150\cdot DM \cdot(\frac{1}{{f_1}^2}-\frac{1}{{f_2}^2})}
\end{equation}
between the highest ($f_2=327.50$ MHz) and lowest frequency IF ($f_1=321.56$ MHz) using the times of steepest rise for four of the bursts. The delay we found was $-0.94\pm3.65$s corresponding to a DM of $-653\pm2530$ $\mathrm{cm^{-3}}$pc, consistent with the value above, but a very weak constraint.\\
From the poorer quality of the exponential fits relative to the linear fits we are inclined to conclude that the shape of the tails of the observed bursts are dominated by tails in the intrinsic emission. It seems justified that the average decay time from the exponential fits (0.72 min) is merely an upper limit for the true scattering time. In general we can state that for scattering times corresponding to distances not far beyond the Galactic Center the intrinsic burst shape will not differ greatly from the observed burst shape, besides any unresolved variability on very short timescales. The reason for this is that the duration of the observed bursts is much longer ($\simeq10$ min) than any reasonable scattering time for sources near the GC.\\
We can work out the original burst profile using theorems for Laplace transforms. The intrinsic emission I(t) is convolved with the scattering function $\zeta(t)$. This gives the observed burst O(t):
\begin{equation}
\mathrm{O=I*\zeta}
\end{equation}
where * denotes convolution. For simple scattering, $\zeta$ is the product of the Heaviside step function $\Pi$ and an exponential:
\begin{equation}
\mathrm{\zeta(t)=\Pi(t)\cdot \exp(\frac{-t}{\tau_{sc}})}
\end{equation}
The Laplace transform of this product is equal to $\mathrm{\frac{1}{s+\alpha}}$, with s the transformed coordinate and $\mathrm{\alpha=1/\tau_{sc}}$. Now, using the theorems for Laplace transforms of convolved functions and derivatives we find:
\begin{equation}
\mathrm{I\cdot \kappa=\alpha \cdot O+\frac{dO}{dt}}
\label{reconstruction}
\end{equation}
with $\kappa$ a constant for normalization. If no emission is absorbed, it follows that $\kappa=\alpha$. Thus, we could reconstruct the intrinsic, unscattered burst from the observed burst if we knew the scattering time $\tau_{sc}$. If the observed burst is represented very accurately by three straight lines for the steep rise, the gradual brightening and the steep decay, the original burst must have the same slopes. It then follows that $\tau_{sc}=1/\alpha=0$, hence no scattering, unless there are faults, i.e. sudden "jumps", in the intensity of the intrinsic emission. So the breaks link scattering times and fault sizes.\\
Without any assumptions on the possible degree of faulting in the intrinsic emission, we can find an upper limit for the scattering times using the end of the tails of the observed bursts. The slopes seem constant until the flux is essentially zero for at least three of the bursts. For the end of the tail of the second burst, which is relatively noisy, this is not so clear. Equation~\ref{reconstruction} then imposes an upper limit on the scattering time $\tau_{sc}$ from the condition that the intrinsic emission cannot be negative. This means that the scattering time must be smaller than the time resolution for which we can determine the slopes with confidence: 10s. This implies that GCRT J1745-3009 cannot be located far beyond the GC. From the NE2001 model of \citet{Cordes} we find a pulse broadening time of 9.93s at 325 MHz for a distance of 11 kpc in the direction of GCRT J1745-3009.\\
We conclude from this discussion that the observed bursts depicted in figure~\ref{allburstsoneplot} will closely resemble the intrinsic bursts. Models will need to explain the asymmetry of the bursts, the steep rise, the more gradual brightening and the steep decay and the breaks between them as well as the fact that the brightest emission is seen just before the steep decay. 

\section{Conclusions}
We have derived new upper limits on the quiescent emission of GCRT J1745-3009 at seven epochs. Six observations were made in the 92 cm band and one in the 21 cm band. The 92 cm observation of GCRT J1745-3009 on 2005 March 24 with the WSRT was the second deepest until that time. Five of these seven epochs constitute the oldest set of 92 cm observations taken of the Galactic Center. The nondetections at those epochs do not provide evidence for the double neutron star binary model \citep{Turolla} with a geodetic precession period close to 18 years. However, geodetic precession times could well be somewhat longer. \\
We have reproduced the lightcurve of the discovery dataset of GCRT J1745-3009 more accurately and more completely than in the discovery paper. We see that the shapes of the five bursts are consistent: a steep rise, a gradual brightening and a steep decay. We have improved the $5\sigma$ upper limit on interburst emission from 75 mJy $\mathrm{beam^{-1}}$ to 31 mJy $\mathrm{beam^{-1}}$. Also, we further constrained the $5\sigma$ upper limit on the fractional circular polarization from $15\%$ to $8\%$. We determined the recurrence interval between bursts more accurately: $77.012\pm0.021$ min. We see no evidence for aperiodicity, but we do find that the duration of the bursts varies at the level of a few $\%$. We derived a very steep spectral index, $\alpha=-6.5\pm3.4$. We have investigated scattering and we have shown that scattering times must be less than 10s. This implies that GCRT J1745-3009 cannot be located far beyond the GC. It also means that the shape of the observed bursts will differ little from the intrinsic emission. Models for GCRT J1745-3009 have to explain the asymmetry in the shape of the bursts and in particular the gradual brightening until the steep decay. Some of the suggested models \citep{Turolla,ZhuXu} predict symmetric bursts. The simplest interpretations of those models can now be ruled out, but it is conceivable that the asymmetry in the bursts could be achieved by adding some complexity to those models. Our results favour a rotating system, like the white dwarf pulsar \citep{ZhangGil}, because that can explain the high level of periodicity we see. We have shown that it is very unlikely that this transient is an incoherent synchrotron emitter, because it would have to be closer than 14 pc, unless the emitting region is moving at a relativistic velocity. Although we now have more contraints on the properties of this source, we are still unsure about its basic model.\\
A better understanding of its nature should come from more detections by long time monitoring with high sensitivity and high angular resolution, to tackle the confusion limit and to reduce the number of possible optical counterparts. The next generation of radio telescopes, like LOFAR \citep[see, e.g.,][]{Fender}, will help to do so. The most pressing issue in revealing the nature of GCRT J1745-3009 is still the determination of its distance, which could be achieved by a new detection with sufficient bandwidth between sidebands, in order to measure the time delay from dispersion towards the Galactic Center.\\
We have also investigated possible transient behaviour of a source on the opposite side of the supernova remnant G359.1-0.5 but we found no compelling evidence for variability.

\begin{acknowledgements}
We thank Tao An from Shanghai Astronomical Observatory for supplying us with an A-configuration model of the radio galaxy 1938-155. This model was necessary to phase calibrate our 1986 March 29 observations. We thank the anonymous referee for encouraging us to reproduce the lightcurve from the 2002 discovery dataset of GCRT J1745-3009 after we reported low noise levels in a map from that dataset. The National Radio Astronomy Observatory is a facility of the National Science Foundation operated under cooperative agreement by Associated Universities, Inc.
\end{acknowledgements}

\end{document}